\documentclass[preprint,showpacs,preprintnumbers,amsmath,amssymb]{revtex4}
\usepackage{graphicx}
\usepackage{floatflt}
\usepackage{dcolumn}
\usepackage{bm}
\usepackage{rotating}
\usepackage{natbib}
\usepackage{txfonts}

\begin{document}
 \title{ Asymmetric nuclear matter and neutron-skin in \\ extended
relativistic mean field model}

   \author{B. K. Agrawal}
              \email{bijay.agrawal@saha.ac.in}
   \affiliation{Saha Institute of Nuclear Physics, Kolkata - 700064, India.}

\date{\today}

\preprint{APS/123-QED}

\begin{abstract}

The density dependence of the symmetry energy,  instrumental in
understanding the behaviour of the asymmetric nuclear matter, is
investigated within the extended relativistic mean field (ERMF) model
which includes the contributions from  the self and  mixed interaction
terms for the scalar-isoscalar ($\sigma$), vector-isoscalar ($\omega$)
and vector-isovector ($\rho$) mesons upto the quartic order.  Each of the
26 different parameterizations of the ERMF model employed are compatible
with the bulk properties of the finite nuclei.  The behaviour of the
symmetry energy for several parameter sets are found to be consistent
with the empirical constraints on them as extracted from the analyses of
the isospin diffusion data. The neutron-skin thickness in the $^{208}$Pb
nucleus for these parameter sets of the ERMF model lie in the range of
$\sim 0.20 - 0.24$ fm which is in harmony with the ones   predicted by
the Skyrme Hartree-Fock model.  We also investigate the role of various
mixed interaction terms which are crucial for the density dependence of
the symmetry energy.

 \end{abstract}
   \maketitle
\section{Introduction}

The accurate knowledge of the equation of state (EOS)  for 
asymmetric nuclear matter is important in understanding the structure
of finite nuclei away from the stability line and critical issues in
astrophysics.  The EOS for asymmetric nuclear matter  is mainly
governed by the density dependence of the nuclear  symmetry energy.
Indeed, significant progress has been made in understanding the
behaviour of the symmetry energy at the subnormal densities from
the analyses  of  the  isospin diffusion data in heavy ion collision
\cite{Tsang04,Chen05a,Li05a,Li08} and from the available data for
the neutron skin thickness of several nuclei \cite{Centelles09}. The
constraints on the density dependence of the nuclear symmetry energy as
extracted from the isospin diffusion data agree with the ones deduced
from the isoscaling analyses of isotope ratios in intermediate energy
heavy-ion collisions \cite{Shetty07}.  The experimental data on the
isotopic dependence of the nuclear giant monopole resonance  in even-A
Sn isotopes \cite{Garg07,Li07} also provides some informations on
the nuclear symmetry energy which is in agreement with those derived
from the analyses of the isospin diffusion data.  The behaviour of
nuclear symmetry energy at supranormal densities is largely unknown.
Theoretically, at supranormal  densities, even the issue of whether the
symmetry energy increases or decreases with density still remains unresolved.
The precise measurements of the properties of the compact stars and the
transport model analyses of the heavy-ion collisions at intermediate and
high energies can provide some constraints on the high density behaviour
of the symmetry energy.

The density dependence of symmetry energy obtained using the Skyrme
Hartree-Fock models have been confronted with the constraints extracted
by analyzing the isospin diffusion data \cite{Chen05}.  The symmetry
energy and its slope and the curvature parameter at the saturation
density  for only four out of 21 different parameter sets of the Skyrme
interactions are found to be consistent with the ones extracted from
the isospin diffusion data \cite{Chen05}.  It may be pointed out that
all of these four parameter sets are obtained by fitting the experimental
data for  the binding energies and charge radii for finite nuclei.
Scenario is not the same when the similar investigation \cite{Chen07}
is carried out using three different versions of the relativistic mean
field (RMF) models, namely, (i) models with meson field self interaction
(ii) models with density dependent meson-nucleon couplings and (iii)
point coupling models without meson fields.  Out of 23 parameter sets of
these RMF models, only a few are found to yield symmetry energies
and their density dependence which are consistent with the empirical
constraints imposed by isospin diffusion data.  In particular, 10
different parameter sets were considered for the model (i), but, only two
of them could yield behaviour of the symmetry energy consistent with the
empirical constraints.  Both of these parameter sets are obtained using
nuclear matter observables, instead of fit to  the bulk properties of
finite nuclei. The density dependence of the symmetry energy  is studied
also for several other parameter sets of the RMF model \cite{Vidana09}.
None of these parameter sets yield acceptable results for the density
dependence of the symmetry energy.  Very recently \cite{Sulaksono09},
density dependence of the symmetry energy has been studied using extended
relativistic mean-field (ERMF) model which includes the contributions from
self and mixed interaction terms for the $\sigma$, $\omega$ and $\rho$
mesons  upto the quartic order.  Even in the ERMF model, the density
dependence of the symmetry energy obtained for several parameter sets
are found to be inadequate.  One of the existing parameter set is fine
tuned in Ref.  \cite{Sulaksono09} so that the resulting behaviour of the
symmetry energy can fulfill the empirical constraints on them. However,
this parameter set is not capable of reproducing the experimental data on
the bulk properties of finite nuclei.  We would like to emphasize that the
ERMF model, due to the presence of the various mixed interaction terms,
can yield wide variations in the density dependence of the symmetry
energy without affecting the quality of the fit to the bulk properties
of finite nuclei \cite{Furnstahl02,Sil05}.  Nevertheless, the slope
of the symmetry energy at the saturation density or alternatively the
neutron-skin thickness for the parameter sets of the ERMF model considered
in Ref. \cite{Sulaksono09} are either too low or quite high.

In the present work we investigate the density dependence of the symmetry
energy using 26 different parameterizations of the ERMF model.  All the
parameterizations of the ERMF model considered are obtained by fitting
the experimental data on the binding energy and charge radius for the
finite nuclei.  Furthermore, these parameter sets yield the neutron-skin
thickness in $^{208}$Pb nucleus  which vary over  a wide range from
$0.16 - 0.28$ fm.  We find that quite a few of these parameterizations
can fulfill the empirical constraints  on the symmetry energy. Role of
various mixed interaction terms are also investigated.

The paper is organized as follows. In Sec. II we describe the ERMF model
in brief. In Sec. III, we provide the expressions used to compute various
quantities associated with the nuclear matter along with the empirical
constraints on them.  In Sec. IV, the results obtained using different
parameterizations of the ERMF model are confronted  with the empirical
constraints on the symmetry energy as extracted from the analyses of
the isospin diffusion data. The role of mixed interaction terms of the
ERMF model  which are important in determining the variations in the
density dependence of the symmetry energy is investigated in Sec. V.
In Sec. VI we state our conclusions.

\section {Extended  Relativistic Mean Field Model}
\label{sec:model}

The ERMF model 
includes the contributions from  the self and  mixed interaction
terms for the scalar-isoscalar ($\sigma$), vector-isoscalar ($\omega$)
and vector-isovector ($\rho$) mesons upto the quartic order. 
Mixed interaction terms
involving $\rho$-meson  field enables one to vary the density dependence
of the symmetry energy coefficient and the neutron skin thickness in
heavy nuclei  over a wide range without affecting the other properties
of finite nuclei \cite{Furnstahl02,Sil05}.  The contribution from the
self interaction of $\omega$-mesons plays important role in varying the
high density behaviour of the EOS and also prevents instabilities in the
calculation of the EOS \cite{Sugahara94,Muller96}.  On the other hand
expectation value of the $\rho$-meson field is order of magnitude smaller
than that for the $\omega$-meson  field \cite{Serot97}. Thus, inclusion
of the $\rho$-meson self interaction can affect the properties of the
finite nuclei and neutron stars  only very marginally \cite{Muller96}.
The effective Lagrangian density
for the ERMF model can be written as,
\begin{equation}
\label{eq:lden}
{\cal L}= {\cal L_{NM}}+{\cal L_{\sigma}} + {\cal L_{\omega}} + {\cal
L_{\mathbf{\rho}}} + {\cal L_{\sigma\omega\mathbf{\rho}}}. 
\end{equation}
where the nucleonic and mesonic Lagrangian ${\cal L_{NM}}$ can be written as, 
\begin{equation}
\label{eq:lbm}
{\cal L_{NM}} = \sum_{J=n,p} \overline{\Psi}_{J}[i\gamma^{\mu}\partial_{\mu}-(M-g_{\sigma} \sigma)-(g_{\omega }\gamma^{\mu} \omega_{\mu}+\frac{1}{2}g_{\mathbf{\rho}}\gamma^{\mu}\tau .\mathbf{\rho}_{\mu})]\Psi. 
\end{equation}
Here, the sum is taken over the neutrons and protons.
$\tau$ are the isospin matrices. The Lagrangian describing
self interactions for $\sigma$, $\omega$,   and $\rho$ mesons can be
written as,
\begin{equation}
\label{eq:lsig}
{\cal L_{\sigma}} =
\frac{1}{2}(\partial_{\mu}\sigma\partial^{\mu}\sigma-m_{\sigma}^2\sigma^2)
-\frac{{\kappa_3}}{6M}
g_{\sigma}m_{\sigma}^2\sigma^3-\frac{{\kappa_4}}{24M^2}g_{\sigma}^2 m_{\sigma}^2\sigma^4,
\end{equation}
\begin{equation}
\label{eq:lome}
{\cal L_{\omega}} =
-\frac{1}{4}\omega_{\mu\nu}\omega^{\mu\nu}+\frac{1}{2}m_{\omega}^2\omega_{\mu}\omega^{\mu}+\frac{1}{24}\zeta_0 g_{\omega}^{2}(\omega_{\mu}\omega^{\mu})^{2},
\end{equation}
\begin{equation}
\label{eq:lrho}
{\cal L_{\mathbf{\rho}}} =
-\frac{1}{4}\mathbf{\rho}_{\mu\nu}\mathbf{\rho}^{\mu\nu}+\frac{1}{2}m_{\rho}^2\mathbf{\rho}_{\mu}\mathbf{\rho}^{\mu}.
\end{equation}
The $\omega^{\mu\nu}$, $\mathbf{\rho}^{\mu\nu}$ are field tensors corresponding to the 
$\omega$ and $\rho$ mesons, and can be defined as 
$\omega^{\mu\nu}=\partial^{\mu}\omega^{\nu}-\partial^{\nu}\omega^{\mu}$ and 
$\mathbf{\rho}^{\mu\nu}=\partial^{\mu}\mathbf{\rho}^{\nu}-\partial^{\nu}\mathbf{\rho}^{\mu}$. 
The mixed interactions of $\sigma, \omega$, and $\mathbf{\rho}$ mesons ${\cal L_{\sigma\omega\rho}}$ can be written as, 
\begin{equation}
\label{eq:lnon-lin}
\begin{split}
{\cal L_{\sigma\omega\rho}} & =
\frac{\eta_1}{2M}g_{\sigma}m_{\omega}^2\sigma\omega_{\mu}\omega^{\mu}+ 
\frac{\eta_2}{4M^2}g_{\sigma}^2 m_{\omega}^2\sigma^2\omega_{\mu}\omega^{\mu}
+\frac{\eta_{\rho}}{2M}g_{\sigma}m_{\rho }^{2}\sigma\rho_{\mu}\rho^{\mu}
\\
&+\frac{\eta_{1\rho}}{4M^2}g_{\sigma}^2m_{\rho }^{2}\sigma^2\rho_{\mu}\rho^{\mu}
+\frac{\eta_{2\rho}}{4M^2}g_{\omega}^2m_{\rho
}^{2}\omega_{\mu}\omega^{\mu}\rho_{\mu}\rho^{\mu}.
\end{split}
\end{equation}
The ${\cal L}_{em}$ is Lagrangian for electromagnetic interactions and can be expressed as,
\begin{equation}
\label{eq:lem}
{\cal L}_{em}= -\frac{1}{4}F_{\mu\nu}F^{\mu\nu}- e\overline{\Psi} _{p}\gamma_{\mu}A_{\mu}\Psi_{p},
\end{equation}
where, $A$ is the photon filed and $F^{\mu\nu}=\partial^{\mu}A^{\nu}-\partial^{\nu}A^{\mu}$.
The equation of motion for nucleons, mesons and photons can be derived
from the Lagrangian density defined in Eq.(\ref{eq:lden}). 
The contributions from Eq. (\ref{eq:lem}) are included only for the case
of finite nuclei.

\section{Empirical constraints on symmetry energy}
\label{sec2}

The symmetry energy $E_{\rm sym}$, slope $L$
and curvature $K_{\rm sym}$ can be evaluated as, 

  \begin{eqnarray}
E_{\rm sym}(\rho)=\frac{1}{2}\left .\frac{d^2E(\rho,\delta)}{d\delta^2}\right
|_{\delta=0},\\
L=3\rho_0\left .\frac{d E_{\rm sym}(\rho)}{d\rho}\right |_{\rho=\rho_0},\\
K_{\rm sym}=9\rho^2_0\left .\frac{d^2 E_{\rm sym}(\rho)}{d\rho^2}\right
|_{\rho=\rho_0},\\
\end{eqnarray}
where, $\rho_0$ is the saturation density,  $E(\rho,\delta)$ is the energy
per nucleon at a given density $\rho$ and asymmetry $\delta=(\rho_n -
\rho_p)/\rho$.  The density dependence of the symmetry energy can also  be
expressed in terms of $E_{\rm sym}(\rho)$, $L$ and $K_{\rm sym}$ as,
\begin{equation}
\label{eq:esym}
E_{\rm sym}(\rho) = E_{\rm sym}(\rho_0)+L\left
(\frac{\rho-\rho_0}{3\rho_0}\right )+\frac{K_{\rm
sym}}{2\!}\left
(\frac{\rho-\rho_0}{3\rho_0}\right)^2.
\end{equation}
The above equation represent very well the behaviour of the symmetry
energy at subnormal densities.  At supranormal densities, one needs
to include in Eq. (\ref{eq:esym}) the contributions of higher order terms
\cite{Chen09}. We also evaluate,
 \begin{eqnarray}
\label{eq:k0}
K_0=9\rho^2_0\left .\frac{d^2 E_0(\rho)}{d\rho^2}\right
|_{\rho=\rho_0},\\
J_0=27\rho^3_0\left .\frac{d^3 E_0(\rho)}{d\rho^3}\right
|_{\rho=\rho_0},\\
\label{eq:ksat2}
K_{\rm sat,2}=K_{\rm asy}-\frac{J_0}{K_0}L,\\
K_{\rm asy}=K_{\rm sym}-6L.
\end{eqnarray}
The $E_0(\rho)=E(\rho,\delta=0)$ is the energy per nucleon for symmetric
nuclear matter.  The $K_0$ is the incompressibility coefficient  of the symmetric
nuclear matter at the saturation density which together with $K_{\rm
sat,2}$ can yield the value of incompressibility coefficient  for asymmetric
nuclear matter \cite{Chen09}.  The constraints on the values of
$E_{\rm sym}$, $L$, $K_{\rm asy}$  \cite{Tsang04,Chen05a,Li05a},  $K_0$
\cite{Youngblood99,Lui04,Ma02,Vretenar03,Colo04a,Shlomo06,Li07,Garg07},
$K_{\rm sat,2}$ \cite{Chen09} are,
 \begin{eqnarray}
E_{\rm sym} = 30\pm 5 \text{MeV}\nonumber\\
L = 88\pm 25 \text{MeV}\nonumber\\
K_{\rm asy} = -500\pm 50 \text{MeV}\nonumber\\
K_0 = 240\pm 20 \text{MeV}\nonumber\\
K_{\rm sat,2} = -370\pm 120 \text{MeV}.
\label{eq:const}
\end{eqnarray}

\section{Asymmetric nuclear matter and neutron-skin }

We study the properties of asymmetric nuclear matter for 26 different
parameterizations of the ERMF model.  Each of these parameterizations
are obtained by fitting the experimental data on the binding energies
and charge radii for few closed shell  nuclei.  Twenty one of these
parameter sets were obtained in our earlier work \cite{Dhiman07} which
correspond to different combinations of neutron-skin thickness $\Delta R$
in $^{208}$Pb nucleus and  the $\omega$ -meson self-coupling strength
$\zeta_0=\zeta g_\omega^2$ (Eq. \ref{eq:lome}).  Hereafter, we name these 21 parameter sets
as BSR1 - BSR21  according to  the values of $\zeta$ and $\Delta R$.
The parameter sets BSR1, BSR2, ...,BSR7 correspond to $\Delta R = 0.16,
0.18, ...., 0.28$ fm with $\zeta=0$.  Similarly, the parameter sets BSR8,
BSR9, ..., BSR14 and BSR15, BSR16, ...,BSR21 correspond to $\Delta R =
0.16, 0.18, ...,0.28$ fm, but, with $\zeta = 0.03\text{ and } 0.06$,
respectively.  The set of experimental data for the binding energies
and charge radii used to generate the parameter sets BSR1 - BSR21 are
exactly the same. It can be seen from Ref. \cite{Dhiman07} that the rms
error on the total binding energy and the charge radius for all our
21 parameter sets are more or less the same.  The rms errors for the
total binding energy are 1.5 - 1.8 MeV and that for the charge
radii lie within the  $0.025 - 0.04$ fm.  Other five parameter sets
considered are, FSUGold \cite{Todd-Rutel05}, FSUGZ03 \cite{Kumar06},
G2 \cite{Furnstahl97}, TM1 \cite{Sugahara94} and TM1$^*$ \cite{Estal01}.

\begin{figure}
\resizebox{6.5in}{!}{ \includegraphics[]{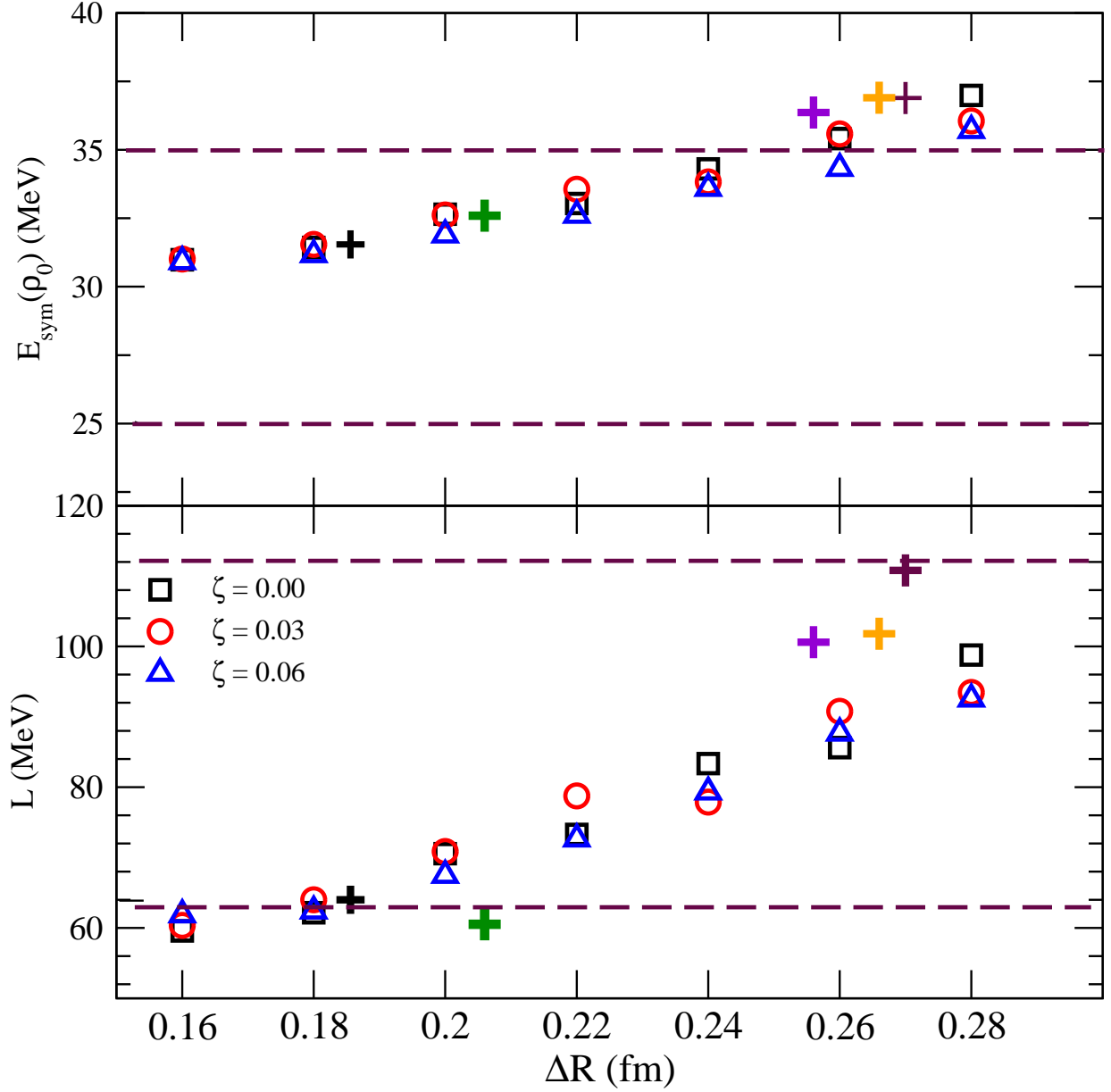}}
\caption{\label{fig1} (colour online)
The symmetry energy $E_{\rm sym}(\rho_0)$ and its slope $L$  plotted
against the neutron-skin thickness $\Delta R$ in the $^{208}$Pb nucleus
for 26 different parameterizations of the ERMF model (see also Table
\ref{tab1}).  The open squares, circles and triangles represent the
results for the parameter sets BSR1 - BSR7, BSR8 - BSR14 and BSR15 -
BSR21, respectively.  The symbol plus with different colours depict
the results for the FSUGold (green), FSUGZ03 (black), G2 (purple), TM1
(maroon) and TM1$^*$ (orange). The dashed lines represent the constraint
on $E_{\rm sym}(\rho_0)$ and $L$ as given in Eq. (\ref{eq:const}).
}
  \end{figure}

\begin{figure}
\resizebox{6.5in}{!}{ \includegraphics[]{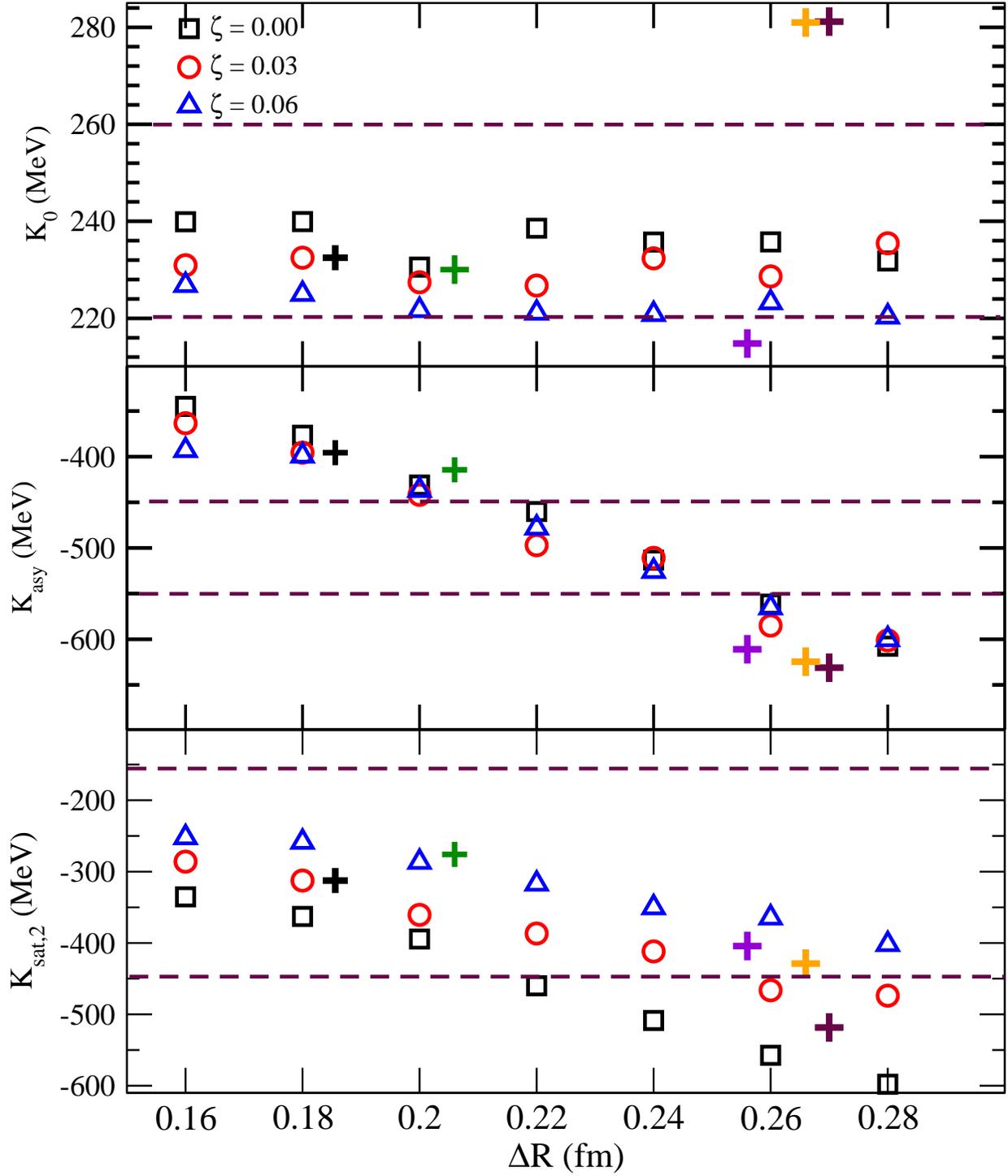}}
\caption{\label{fig2} (colour online) Same as Fig. \ref{fig1}, but,
for $K_0$, $K_{\rm asy}$ and $K_{\rm sat,2}$ } 
  \end{figure}

The density dependence of the symmetry energy $E_{\rm sym}(\rho)$
plays central role in understanding the behaviour of the asymmetric
nuclear matter. The $E_{\rm sym}(\rho)$ at subnormal densities
can be expressed in terms of the $E_{\rm sym}(\rho_0)$, slope $L$
and the curvature $K_{\rm sym}$ or $K_{\rm asy}$ as given by
Eq. (\ref{eq:esym}). We mainly focus on the values of $E_{\rm
sym}(\rho_0)$, $L$ and $K_{\rm asy}$ as obtained for different
parameterizations of the ERMF model. In addition, we also calculate
the curvature parameter $K_{\rm sat,2}$ (Eq.  \ref{eq:ksat2}) which
together with $K_0$  yields the values for the incompressibility
coefficient for the asymmetric nuclear matter \cite{Chen09}.  In Table
\ref{tab1} we list the values of the $\Delta R$, $B/A$, $\rho_0$, $K_0$,
$E_{\rm sym}$, $L$, $K_{\rm sym}$, $K_{\rm asy}$  and $K_{\rm sat,2}$
for different parameterizations of ERMF model.  For the comparison, in
the last row of Table \ref{tab1}, we also provide the results obtained
for most commonly used NL3  parameterization \cite{Lalazissis97} of
the conventional RMF model which includes the non-linear terms only
for the $\sigma$ meson.  For better insight, in Fig. \ref{fig1},  the
values of $E_{\rm sym}(\rho_0)$ and $L$ are plotted  against $\Delta R$
and $\zeta$ for the parameter sets BSR1 - BSR7 (squares), BSR8 - BSR14
(circles) and BSR15 - BSR21 (triangles).  Similar plots for $K_0$,
$K_{\rm asy}$ and $K_{\rm sat,2}$ are displayed in Fig. \ref{fig2}.
The results for the other parameter sets FSUGold, FSUGZ03, G2, TM1 and
TM1$^*$ are also depicted in Figs. \ref{fig1} and \ref{fig2}.  We first
focus on the results obtained for our 21 parameter sets BSR1 - BSR21. It
can be seen from Figs. \ref{fig1} and \ref{fig2} that the values of the
isovector quantities $E_{\rm sym}(\rho_0)$, $L$ and $K_{\rm asy}$ depends
mainly on the value of $\Delta R$.  The isoscalar quantity $K_0$ is more
sensitive to the choice of $\zeta$.  The values of $K_{\rm sat,2}$,
however, depends on both the $\zeta$ and $\Delta R$.  The constraints
(Eq. \ref{eq:const}) on the values of $E_{\rm sym}(\rho_0)$ and $L$
are satisfied by our parameter sets for which $\Delta R = 0.18 -
0.24$ fm.  The  constraint on the  values of $K_0$ is satisfied by
all the 21 parameter sets BSR1 - BSR21.  The constraints on the value
of $K_{\rm asy}$ are satisfied by our parameterizations with $\Delta
R = 0.22 - 0.24$ fm.  On the other hand, except for three parameter
sets, all of our other parameterizations satisfy the constraint on
the value of $K_{\rm sat,2}$.  These three parameter sets are BSR5,
BSR6 and BSR7 which correspond to $\zeta = 0$ with $\Delta R = 0.24,
0.26 \text{ and } 0.28$ fm, respectively.  Other parameter sets FSUGold,
FSUGZ03, G2, TM1 and TM1$^*$ are  not compatible simultaneously  with
the various constraints summarized in Eq. (\ref{eq:const}).  In short,
it appears that only five parameter sets BSR4, BSR11, BSR12, BSR18 and
BSR19 with $\Delta R = 0.22 - 0.24$ fm obey the empirical constraints
of Eq. (\ref{eq:const}) very well.  Our parameter sets BSR3, BSR10 and
BSR17 with $\Delta R = 0.20$ fm also satisfy all the constraints of
Eq. (\ref{eq:const}) except for the $K_{\rm asy}$. The values of $K_{\rm
asy}$  for these parameter sets are only marginally away from the ones
given by Eq. (\ref{eq:const}). So, on the basis of the these constraints,
the ERMF model  predicts the value of neutron-skin thickness in the
$^{208}$Pb nucleus to be $\sim 0.20 - 0.24$ fm.  Similar investigations
using the Skyrme Hartree-Fock models \cite{Chen05} predicted $\Delta R =
0.18 - 0.26$ fm.  Thus, the empirical constraints extracted from the
isospin diffusion data predict more or less model independent values
for the neutron-skin thickness.  In Ref. \cite{Sulaksono09}, various
parameter sets of the ERMF model considered are the FSUGZ00, FSUGZ03,
FSUGZ06 and G2. For these cases, $\Delta R$ is either $\sim 0.19$ or
$\sim 0.26$ fm (see also Table \ref{tab1}).  These values of $\Delta R$
are seem to be either little smaller or quite larger.  Consequently,
none of the parameter sets considered in Ref. \cite{Sulaksono09} are
consistent with all the constraints of Eq. (\ref{eq:const}).

\begin{figure}
\resizebox{6.5in}{!}{ \includegraphics[]{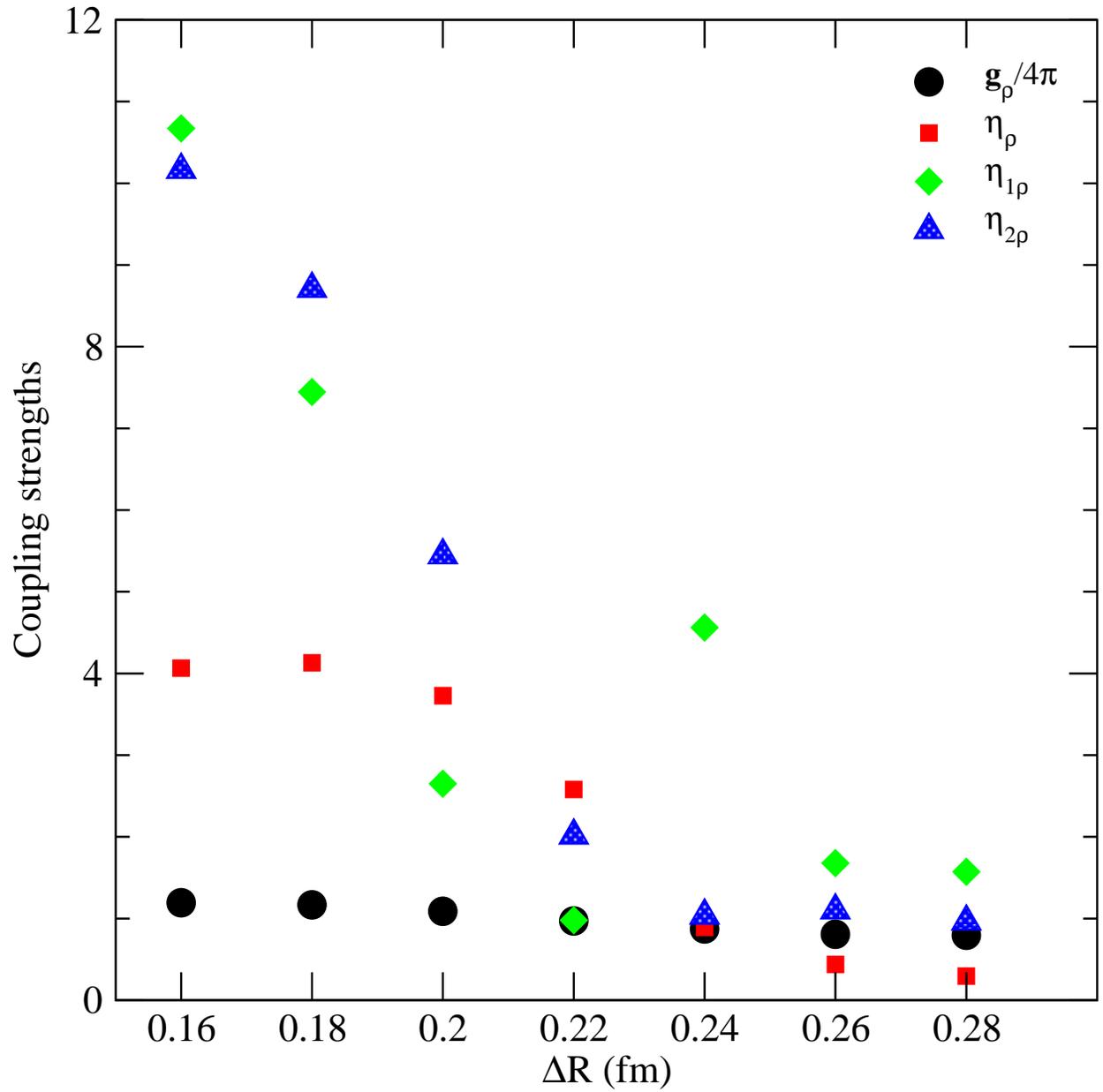}}
\caption{\label{fig3} (colour online)
The coupling strengths $g_\rho/4\pi$, $\eta_\rho$, $\eta_{1\rho}$ and
$\eta_{2\rho}$ for the parameter sets BSR8 - BSR14 plotted against
$\Delta R$. 
 }
\end{figure}

The symmetry energy and its density dependence in the ERMF model
is mainly governed by values of the coupling strengths  $g_\rho$,
$\eta_\rho$, $\eta_{1\rho}$ and $\eta_{2\rho}$ (Eqs. \ref{eq:lbm}
and \ref{eq:lnon-lin}).  The strengths $\eta_\rho$, $\eta_{1\rho}$ and
$\eta_{2\rho}$ determines the contributions of the mixed interaction
terms which account for  the coupling of the isoscalar $\sigma$ and
$\omega$ mesons to the isovector $\rho$ mesons. It may be emphasized,
in the conventional RMF models, the contributions of these mixed
interaction terms are ignored (i.e., $\eta_\rho$= $\eta_{1\rho}$=
$\eta_{2\rho} =0$). We have used our parameter sets BSR1 - BSR21 to
look into the variations of the $g_\rho$, $\eta_\rho$, $\eta_{1\rho}$
and $\eta_{2\rho}$ with $\Delta R$. As an illustration,
in Fig. \ref{fig3}, we plot the values of the $g_\rho/4\pi$, $\eta_\rho$,
$\eta_{1\rho}$ and $\eta_{2\rho}$ for the parameter sets BSR8 $-$ BSR14
which correspond to different values of $\Delta R$ with $\zeta =0.03$.
Scenario for the parameter sets BSR1- BSR7 and BSR15- BSR21 (not
shown here) is analogous to that of BSR8 $-$ BSR14. We
can see from Fig. \ref{fig3}, there is an overall decrease in the
values of these coupling strengths with increase in $\Delta R$.  It is
interesting to note that all the coupling strengths are nearly unity
for $\Delta R \sim 0.22$ fm. Coincidently, for $\Delta R \sim 0.22$
fm, the empirical constraints on the behaviour of the symmetry energy
as extracted from the isospin diffusion data are also satisfied very
well. The  strengths $\eta_\rho$, $\eta_{1\rho}$ and $\eta_{2\rho}$
tend to vanish for higher values of $\Delta R$. In other words, if the
contributions of the mixed interaction terms in question are ignored,
as in the case of the conventional RMF model, the parameters obtained
by fit to the experimental data on the binding energy and charge radius
would give rise to $\Delta R \gtrsim 0.26$ fm.  The values of $\Delta
R\sim 0.22$ fm, as favoured by the isospin diffusion data, can be achieved
within the conventional RMF model only at the expense of the quality of
fit to the bulk properties of the finite nuclei.

\section {Role of mixed interactions }

We would like to investigate the role of various mixed interaction
terms which are crucial in determining the density dependence of
the symmetry energy. In particular, we investigate the effects of
the terms which  account for  the coupling of the isoscalar $\sigma$ and
$\omega$ mesons to the isovector $\rho$ mesons. The coupling constants
for these terms are $\eta_\rho$, $\eta_{1\rho}$ or $\eta_{2\rho}$
(Eq. \ref{eq:lnon-lin}).  The term with coupling constant
$\eta_\rho$ is of the cubic order in the meson fields. Whereas, the
terms with $\eta_{1\rho}$ and $\eta_{2\rho}$ are of quartic order  in
the meson fields. Our objective is to delineate the effects of these
cubic and quartic order mixed interaction terms.  For this purpose, we
generate two different families of interactions.  For the first family
of interactions F1, we put $\eta_{1\rho}=\eta_{2\rho} = 0$ and fit the
remaining coupling constants using appropriate set of experimental data
for the bulk properties of finite nuclei. But, the coupling constants
$\eta_{1\rho}$ and $\eta_{2\rho}$ are also included in the fit for the
second family F2.  Thus, the behaviour of the symmetry energy for the F1
family is governed by the coupling constants $g_\rho$ and $\eta_\rho$ only.
In case of the F2 family, the behaviour of the symmetry energy depends
additionally  on $\eta_{1\rho}$ and $\eta_{2\rho}$.  For both the families
we obtain the parameter sets corresponding to different values of
$\Delta R$ in the range of $0.18 - 0.26$ fm with fixed $\zeta = 0.03$.
The procedure for calibrating the parameters of the model  as well as
the set of experimental data for the binding energies and charge radii
used here are exactly the same as in Ref.  \cite{Kumar06}. In addition,
the parameters are also subjected to the empirical constraints of
Eq. (\ref{eq:const}).

\begin{figure}
\resizebox{6.5in}{!}{ \includegraphics[]{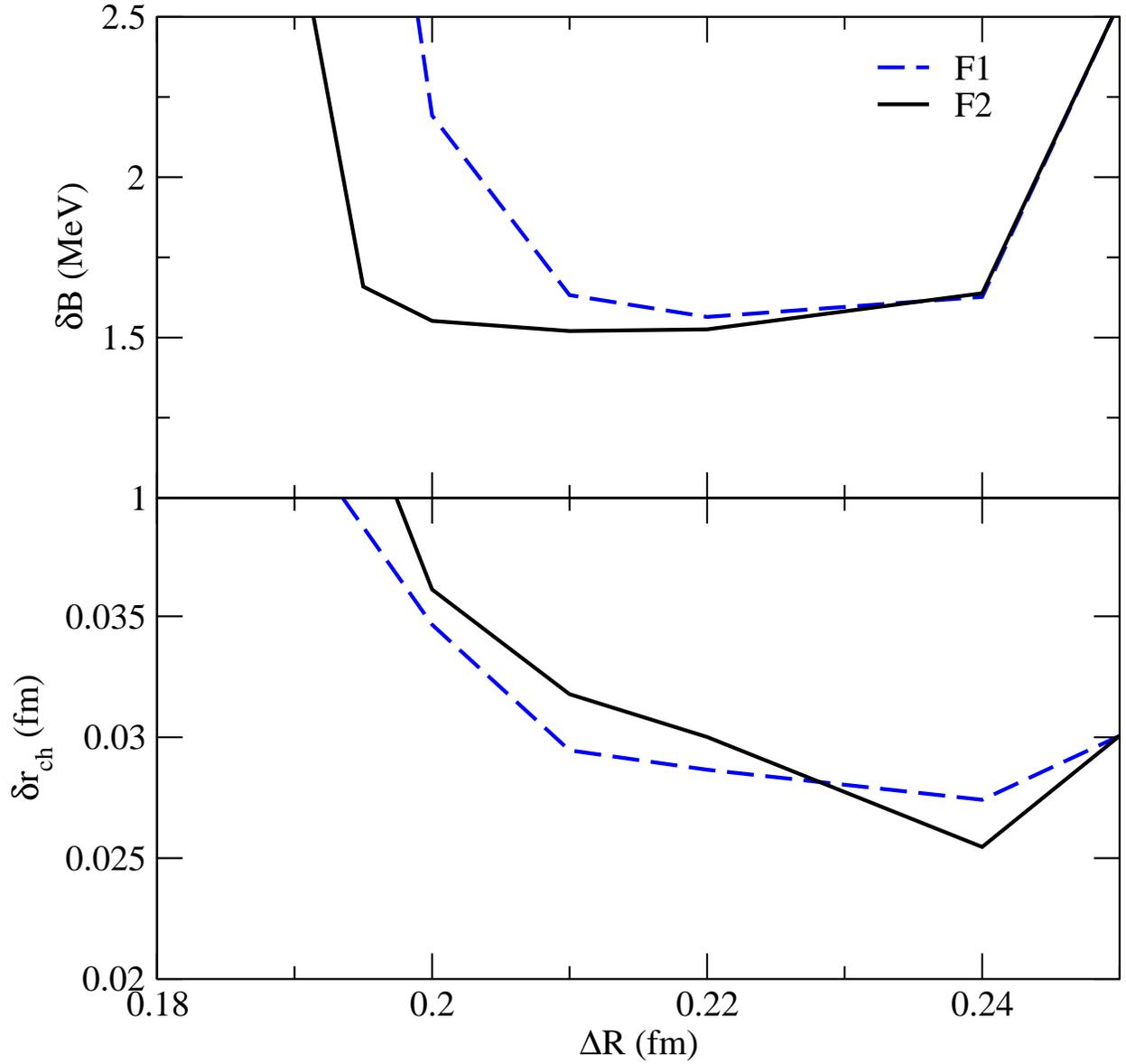}}
\caption{\label{fig4}(colour online)
Plots for the rms errors on the total binding energy $\delta B$ (upper
panel) and the charge radius $\delta r_{\rm ch}$ (lower panel) for the
F1 and F2 families of the interactions.}
\end{figure}

In Fig. \ref{fig4} we plot our results  for the rms errors on the
total binding energies and charge radii.  We see that quality of the
fits to the total binding energies and charge radii are more or less
the same for both the F1 and F2 families of interactions, except for
$\Delta R \leqslant 0.2$ fm.  Therefore, it appears that the quartic
order mixed interaction terms with coupling constants $\eta_{1\rho}$
and $\eta_{2\rho}$ are redundant.  More precisely, one might say that
the values of $\eta_{1\rho}$ and $\eta_{2\rho}$ can not be appropriately
determined by the bulk properties of the finite nuclei.  These quartic
order terms might play important role in fixing the behaviour  of the
symmetry energy at supranormal densities  which is largely unknown
at present. It  can be easily concluded from Fig. \ref{fig4}
that the most preferred value for the neutron-skin thickness in the
$^{208}$Pb nucleus within the ERMF model is $\Delta R\sim 0.22$.
In Table \ref{tab2}, we give the parameter sets for the F1 family of
interactions  obtained for different values of $\Delta R$. These
parameter sets are named  as BKA20, BKA22 and BKA24 which correspond
to $\Delta R = 0.20, 0.22 \text{ and } 0.24$ fm, respectively. In Table
\ref{tab3}, we present the results for the various quantities associated
with the symmetric and asymmetric nuclear matter calculated at the
saturation density using the parameter sets BKA20, BKA22 and BKA24.
For the sake of completeness, we have repeated our calculation for
$\Delta R = 0.22$ fm with $\eta_\rho = \eta_{1\rho}=\eta_{2\rho} = 0$.
In this case, the rms errors $\delta B = 4.8$ MeV and $\delta r_{\rm ch}
= 0.05$ fm are significantly higher compared to the ones obtained for
the F1 and F2 families.  Thus, the  contributions of the mixed interaction
terms seem indispensable in order to satisfy simultaneously the empirical
constraints on the density dependence of the symmetry energy as well as
the experimental data on the bulk properties of the finite nuclei.

 \section{Conclusions}

The density dependence of symmetry energy and the incompressibility
coefficient for the asymmetric nuclear matter are studied using 26
different parameterizations of the ERMF model. The model includes the
contributions from self and mixed interaction  terms for $\sigma$,
$\omega$ and $\rho$ mesons upto the quartic order.  Each of  the
parameterizations considered are compatible with the bulk properties
of the finite nuclei.  Furthermore, these parameter sets yield the
neutron-skin thickness in $^{208}$Pb nucleus  which vary over  a wide
range from $0.16 - 0.28$ fm.  The behaviour of symmetry energy at
subnormal densities for several parameterizations of the ERMF model
corresponding to $\Delta R\sim 0.20 - 0.24$ fm are found to be  consistent
with the empirical constraints on them as extracted from the analyses
of the isospin diffusion data.  The ERMF model prediction for $\Delta R
\sim 0.20 - 0.24$ fm is in reasonable agreement with the ones obtained
in the similar way, but, for the Skyrme Hartree-Fock model \cite{Chen05}.
We have investigated the role of the cubic and quartic order  mixed
interaction terms which are crucial for the density dependence of the
symmetry energy. It is utmost important to include the contributions
at least from the cubic order term to incorporate the empirical constraints
on the density dependence of the symmetry energy without affecting
the quality  of the fit to the bulk properties of the finite nuclei.
The mixed interaction terms of the quartic order might be important to
obtain  the appropriate behaviour of the symmetry energy at supranormal
densities which is largely unknown.

\newpage

\begin{table}
\caption{\label{tab1}
Some bulk properties of the nuclear matter at the saturation density
($\rho_0$): binding energy per nucleon ($B/A$), incompressibility
coefficient for symmetric nuclear matter ($K_0$), symmetry energy
($E_{\rm sym}(\rho_0)$), linear density dependence of the symmetry energy ($L$)
and various quantities ($K_{\rm sym}$), ($K_{\rm asy}$) and ($K_{\rm
sat2}$) as given by Eqs. (\ref{eq:k0}-\ref{eq:ksat2}).  The values for
the neutron-skin thickness $\Delta R$ for the $^{208}\text{Pb}$ nucleus are
also listed.  The superscript '$a$' on several parameter sets 
indicate that they satisfy the constraints of Eq. (\ref{eq:const})
very well.  }

  \begin{ruledtabular}
\begin{tabular}{cccccccccc}
Force& $\Delta _R$& $B/A$&$\rho_0$& $K_0$& $E_{\rm sym}(\rho_0)$&$ L$&$ K_{\rm sym}$& $K_{\rm asy}$&
$K_{\rm sat2}$\\
  & (fm)& (MeV)& (fm$^{-3}$)& (MeV)& (MeV)&(MeV)&(MeV)&(MeV)&(MeV)\\
\hline
BSR1& 0.16 & 16.0   & 0.148  &240  & 31.0 &  60&   13 &  -345& -335  \\ 
BSR2& 0.18  &16.0    &0.149 & 240  & 31.4 &  62&   -4 &  -376& -363 \\ 
BSR3& 0.20 & 16.1   & 0.150  &231  & 32.6 &  71&   -8 &  -431& -395  \\ 
BSR4$^a$& 0.22  &16.1    &0.150 & 239  & 33.0 &  73&  -20 &  -460& -460  \\ 
BSR5& 0.24 & 16.1   & 0.151  &236  & 34.3 &  83&  -13 &  -513& -509 \\
BSR6& 0.26  &16.1    &0.149 & 236  & 35.4 &  86&  -48 &  -562& -557 \\
BSR7& 0.28 & 16.2   & 0.149  &232  & 37.0 &  99&  -15 &  -608& -598    \\
BSR8& 0.16  &16.0   & 0.147 & 231  & 31.0 &  60&   -1 &-363& -286  \\
BSR9& 0.18 & 16.1  &  0.147  &233  & 31.6 &  64&  -12 &-396& -313 \\
BSR10& 0.20  &16.1   & 0.147  &228  & 32.6 &  71&  -17 &-442& -361  \\
BSR11$^a$& 0.22 & 16.1    &0.147 & 227  & 33.6 &  79&  -25 &-497& -387  \\
BSR12$^a$& 0.24  &16.1   & 0.147  &232  & 33.8 &  78&  -44 &-511& -412   \\
BSR13& 0.26  &16.1    &0.147 & 229  & 35.6 &  91&  -40 &-585& -466  \\
BSR14& 0.28 & 16.2   & 0.147  &236  & 36.1 &  94&  -41 &-601& -474  \\
BSR15& 0.16  &16.0   & 0.146  &227  & 30.9 &  62&  -22 &-393& -252 \\
BSR16& 0.18 & 16.1    &0.146 & 225  & 31.2 &  62&  -25 &-399& -258  \\
BSR17& 0.20  &16.1   & 0.146  &222  & 31.9 &  68&  -32 &-437& -287  \\
BSR18$^a$& 0.22  &16.1  &  0.146 & 221  & 32.6 &  73&  -42 &-478& -317  \\
BSR19$^a$& 0.24  &16.1    &0.147  &221  & 33.6 &  79&  -50 &-526& -350  \\
BSR20& 0.26  &16.1   & 0.146 & 223  & 34.3 &  88&  -39 &-565& -365  \\
BSR21& 0.28 & 16.1  &  0.145  &220  & 35.7  & 93 & -45& -600& -402  \\
FSUGold & 0.21  &16.3   & 0.148 & 229&   32.5&   60&  -52& -412& -276 \\
FSUGZ03 &  0.19&  16.1  &   0.147 &  233  &  31.6 &   64&   -11&  -396&  -312\\
G2  &0.26&  -16.1&    0.153&  215 &  36.4&  100&   -7& -611  &-404\\
TM1& 0.27 & 16.3 &   0.145&  281&   36.8 & 111  & 34& -632& -518 \\
TM1$^*$&0.27&   16.3&    0.145&  281&   37&  102&  -14& -625& -429 \\
NL3 & 0.28 & 16.3 &   0.148&  272 &  37.4&  118&  100& -608& -700 \\
\end{tabular}
\end{ruledtabular}
\end{table}

\newpage
\begin{table}
\caption{\label{tab2}
Parameter sets for the F1 family of interactions obtained for different
values  of  neutron-skin thickness in the $^{208}$Pb nucleus.  These
parameter sets are named as BKA20, BKA22 and BKA24 which correspond
to the neutron-skin thickness 0.20, 0.22 and 0.24 fm, respectively.
The couplings $\eta_{1\rho}$ and $ \eta_{2\rho}$ are taken to be zero. The
masses for $\omega$ and $\rho$ mesons are $\frac{m_\omega}{M} = 782$
MeV and $\frac{m_\rho}{M} = 770$ MeV with nucleon mass $M = 939$ MeV.
 }
  \begin{ruledtabular}
\begin{tabular}{ccccccccccc} Force &$\frac{g_\sigma}{4\pi}$&
$\frac{g_\omega}{4\pi}$&$\frac{g_\rho}{4\pi}$
&$\kappa_3$&$\kappa_4$&$\zeta_0$&$\eta_1$&$\eta_2$&$\eta_\rho$&
$\frac{m_\sigma}{M}$\\
\hline
BKA20&0.8042&1.0102&0.9812&1.1523&1.9892&4.8344&0.0005&0.0657&3.6164&0.5430\\
BKA22&0.8462&1.1089&1.0302&1.5500&2.13451& 5.8253&0.1555&0.0697&3.9294
&0.5302\\ BKA24&0.8593&1.1463&0.9381&1.7719&3.1064&
6.2247&0.2700&0.1159&2.4900& 0.5248\\ \end{tabular}
  \end{ruledtabular}
\end{table}

\newpage

\begin{table}
\caption{\label{tab3}
Same as Table \ref{tab1}, but, for the forces BKA20, BKA22 and BKA24.
}
\begin{ruledtabular}
\begin{tabular}{cccccccccc}
Force& $\Delta _R$& $B/A$&$\rho_0$& $K_0$& $E_{\rm sym}$&$ L$&$ K_{\rm
sym}$& $K_{\rm asy}$&
$K_{\rm sat2}$\\
  & (fm)& (MeV)& (fm$^{-3}$)& (MeV)& (MeV)&(MeV)&(MeV)&(MeV)&(MeV)\\
\hline
BKA20& 0.20 & 16.1   & 0.146  &240  & 32.3 &  76&   -15 &  -469& -320  \\
BKA22& 0.22& 16.1   & 0.148  &227  & 33.3 &  79&  -9 &  -483& -382  \\
BKA24& 0.24 & 16.1   & 0.148  &228  & 34.3 &  85&  -15 &  -525& -420  \\
\end{tabular}
\end{ruledtabular}
\end{table}

\newpage

 \end{document}